\documentclass[11pt]{article}
\usepackage{lipsum}    
\usepackage{siunitx}  
\usepackage{graphicx}  
\usepackage{float}     
\usepackage{comment}
\usepackage{amsmath}  
\usepackage{amssymb}
\usepackage{xcolor}
\usepackage{geometry} 
\usepackage{authblk} 
\usepackage{multicol}
\usepackage{caption} 
\usepackage{url}
\usepackage{mathtools}
\usepackage[nottoc]{tocbibind}
\usepackage{verbatim}
\usepackage[export]{adjustbox}
\usepackage{titlesec}

\usepackage{titlesec}
\titlespacing*{\section} {0pt}{1ex}{1ex}    
\titlespacing*{\subsection} {0pt}{1ex}{1ex}    

\expandafter\def\expandafter\normalsize\expandafter{%
    \normalsize%
    \setlength\abovedisplayskip{4pt}%
    \setlength\belowdisplayskip{8pt}%
    \setlength\abovedisplayshortskip{-4pt}%
    \setlength\belowdisplayshortskip{5pt}%
}   
\expandafter\def\expandafter\small\expandafter{%
    \normalsize%
    \setlength\abovedisplayskip{-10pt}%
    \setlength\belowdisplayskip{5pt}%
    \setlength\abovedisplayshortskip{-10pt}%
    \setlength\belowdisplayshortskip{0pt}%
}   

\title{\vspace{-2em}Ion Temperature Measurements in the MAST-U Divertor During Steady State Plasmas and ELM Burn Through Phenomena}

\author[1,2]{Y. Damizia}
\author[2]{S. Elmore}
\author[2]{K. Verhaegh}
\author[2]{P. Ryan}
\author[2]{S. Allan}
\author[3]{F.Federici}
\author[1,2]{N.Osborne}
\author[1]{J. W. Bradley}
\author[*]{the MAST-U Team}
\author[**]{the EUROfusion Tokamak Exploitation Team}

\affil[1]{Electrical Engineering and 
Electronics, University of 
Liverpool, Liverpool, L69 3GJ, UK}

\affil[2]{UK Atomic Energy Authority, Culham Centre for Fusion Energy, Abingdon, OX14 3DB, UK
}%

\affil[3]{ 
Oak Ridge National Laboratory, Oak Ridge, Tennessee 37831, USA}

\affil[*]{See the author list of “Overview of physics results from MAST Upgrade towards
core-pedestal-exhaust integration” by J.R. Harrison et al. to be published in Nuclear Fusion Special Issue: Overview and Summary Papers from the 29th Fusion Energy Conference (London, UK, 16-21 October 2023).}

\affil[**]{See the author list of “Overview of the EUROfusion Tokamak Exploitation programme in support of ITER and DEMO” by E. Joffrin Nuclear Fusion 2024 10.10788/1741-4326/ad2be4.}

\date{\vspace{-5ex}} 
\usepackage{hyperref}
\begin{document}

\maketitle
\begin{abstract}
\noindent 

This study presents ion temperature (\(T_i\)) measurements in the MAST-U divertor, using a Retarding Field Energy Analyzer (RFEA).
Steady state measurements were made during an L-Mode plasma with the strike point on the RFEA. Edge Localized Mode (ELM) measurements were made with the strike point swept over the RFEA. The scenarios are characterized by a plasma current (\(I_p\)) of 750 kA, line average electron density (\(n_e\)) between \(1.6 \times 10^{19}\) and \(4.5 \times 10^{19}\,\text{m}^{-3}\), and Neutral Beam Injection (NBI) power ranging from 1.1 MW to 1.6 MW. The ion temperatures, peaking at approximately 10 eV in steady state, were compared with electron temperatures (\(T_e\)) obtained from Langmuir probes (LP) at the same radial positions.  Preliminary findings reveal a \(T_i/T_e\) ratio in the divertor region less than 1 for shot 48008. High temporal resolution measurements captured the dynamics of ELMs Burn Through, providing \(T_i\) data as a radial distance from the probe peaking around 20 eV.
\end{abstract}

\begin{multicols}{2} 

\section{Introduction}

\begin{figure}[H]
\centering
\includegraphics[width=0.48\textwidth]{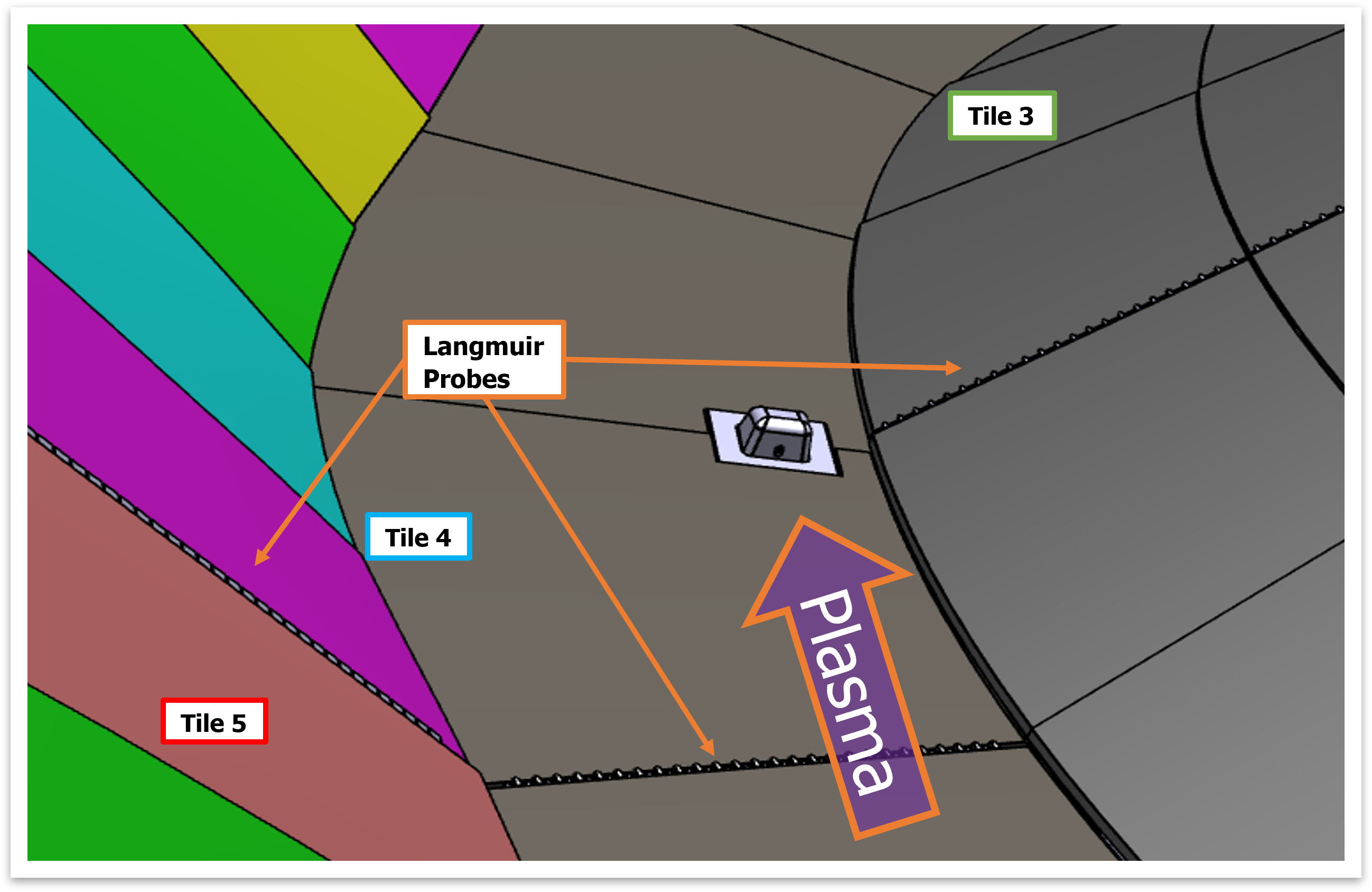}
\caption{CAD of the DSF RFEA location in the MAST-U Divertor.\cite{damizia2024first}}
\label{fig:DSF}
\end{figure}

The Retarding Field Energy Analyzer (RFEA) has proven to be a valuable diagnostic tool \cite{elmore2012divertor, allan2013ion} within MAST-U’s flexible divertor system, specifically integrated into the flat tile of the closed divertor chamber (\autoref{fig:DSF}). This setup allows for the study of various advanced divertor configurations. The RFEA is used to measure the ion temperature (\(T_i\)) by analyzing the energy distribution of ions that pass through a series of biased grids. 

Transient power loading in divertor regions is a significant concern for the operation and longevity of future fusion reactors. The cyclic thermal stresses induced by these transients can lead to accelerated erosion or structural weakening of plasma-facing components (PFCs) \cite{stangeby2000plasma}. Understanding and mitigating these effects are crucial for maintaining material integrity and ensuring the reliability of fusion devices. Accurate measurements of ion temperature (\(T_i\)) are essential for understanding these transient power loads, as they help characterize how energy is distributed and transferred within the divertor during events like ELMs, which are a major source of transient heat flux.

A critical parameter in analyzing plasma behavior is the ion-to-electron temperature ratio (\(T_i/T_e\)). This ratio provides insights into the energy distribution between ions and electrons, impacting transport phenomena and stability within the plasma \cite{ricci2015simulation}. Accurate measurements of \(T_i\) and \(T_e\) are therefore essential for optimizing plasma performance and improving theoretical models.

In this study, the RFEA was employed in the Divertor Science Facility (DSF), a specially designed hole located on tile 4 that allows the insertion of a probe into the divertor region,  to conduct ion temperature (\(T_i\)) measurements under different plasma scenarios, focusing on two distinct shots. The first shot was an L-mode elongated divertor configuration where the plasma strike point was positioned consistently on the RFEA. This setup allowed for measurements of \(T_i\) using the RFEA and electron temperature (\(T_e\)) using Langmuir Probes (LPs) \cite{ryan2023overview} at a different toroidal location.

In the second shot, the plasma strike point was swept over the RFEA multiple times to capture the behavior of Edge Localized Modes (ELMs). During this shot, the RFEA recorded ion temperature data during the ELM events. These ELMs represent bursts of energy that are significant in understanding transient power loads. 
Analyzing the interactions between these temperature profiles and other plasma parameters could provide deeper insights into the transient behaviors in tokamak plasmas. This paper discusses the results obtained from the analysis of this two shots using the RFEA data.

\section{Experimental Setup}
\begin{figure}[H]
\centering
\includegraphics[width=0.45\textwidth]{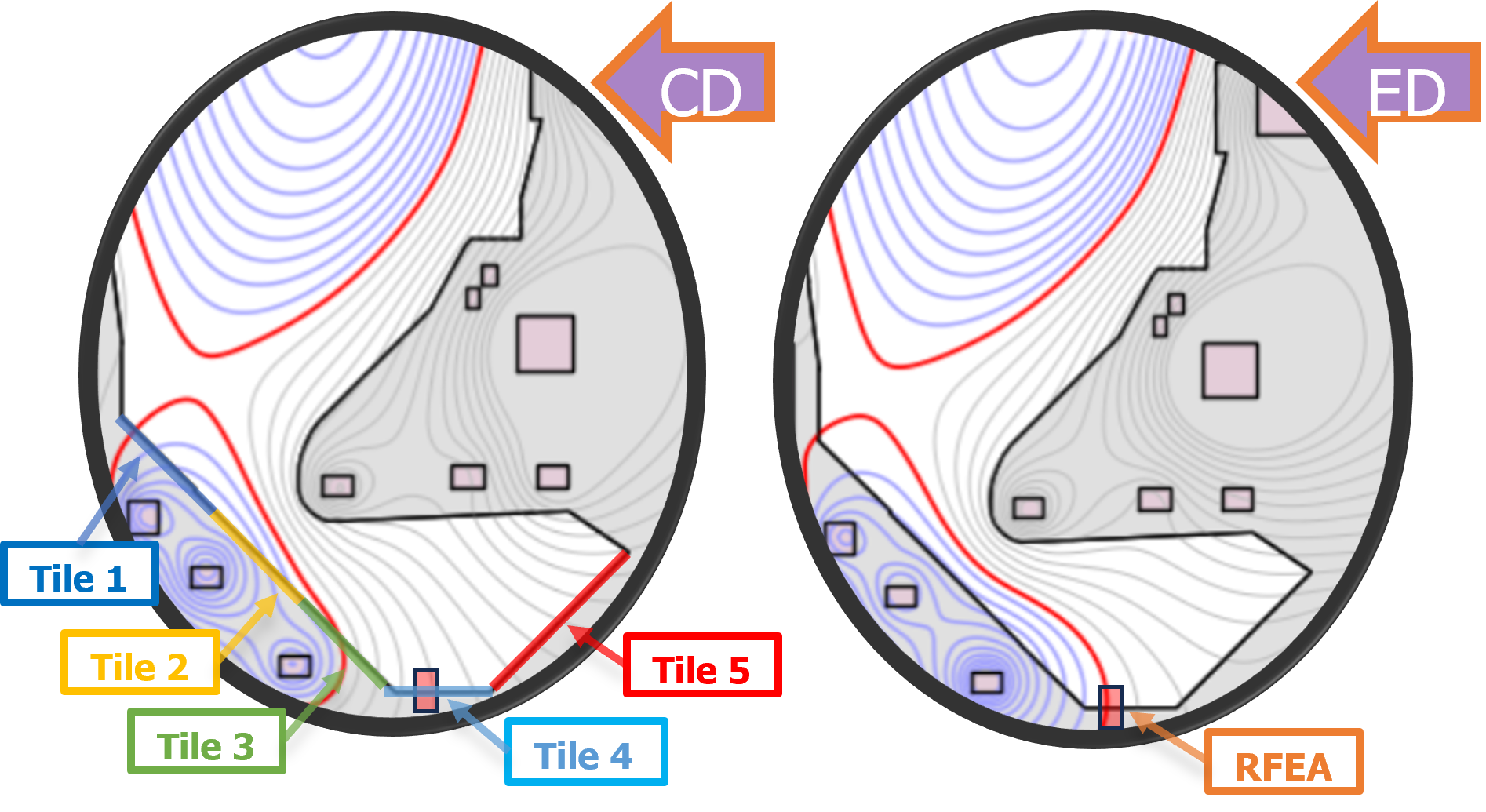}
\caption{Poloidal cross-sectional view of Conventional Divertor (CD) and Elongated Divertor (ED) based on EFIT reconstructions. Divertor tiles (Tile
1–5) and the RFEA location are highlighted.}
\label{fig:ED_configuration}
\end{figure}

The MAST-U Tokamak has been designed with advancements in spherical tokamak technology to improve plasma performance and exhaust management\cite{morris2018mast}. A key feature of MAST-U is its Super-X Divertor chamber, which supports various magnetic configurations. The DSF is located in the horizontal tile 4 of the divertor chamber, and allows for the static insertion of different probe heads into the divertor region\cite{damizia2024first} (see \autoref{fig:DSF}). Measurements can be taken as the plasma strike point reaches the DSF location. As the plasma transitions from a Conventional Divertor (CD) configuration to an Elongated Divertor (ED) as shown in \autoref{fig:ED_configuration}, the RFEA first measures parameters in the Scrape-Off Layer (SOL). Once the strike point traverses the DSF, the RFEA then measures parameters in the Private Flux Region (PFR).
The two shots analyzed in this study are characterized by specific plasma conditions. Shot 47775 maintained a plasma current ($I_{p}$) of 750 kA, with a line average electron density between $2.5 \times 10^{19} \, \text{m}^{-3}$ and $4.5 \times 10^{19} \, \text{m}^{-3}$, and NBI power of approximately \SI{1.1}{\mega\watt}. In contrast, shot 48008 operated with the same plasma current of 750 kA, a line average density range of $1.6 \times 10^{19} \, \text{m}^{-3}$ to $3.7 \times 10^{19} \, \text{m}^{-3}$, and a slightly higher NBI power of \SI{1.6}{\mega\watt} (see \autoref{tab:tokamak_shots}).

\subsection{MAST-U RFEA}

RFEAs are considered to be the most reliable
and least perturbing method of measuring filaments ion temperatures in the scrape-off layer
on different tokamaks  like MAST\cite{allan2016ion}\cite{elmore2012divertor}\cite{elmore2013scrape}\cite{elmore2016scrape}, JET \cite{pitts2003jet}, Tore Supra  \cite{kovcan2008reliability}, and Alcator C \cite{nachtrieb2000omegatron}.

\begin{figure}[H]
\centering
\includegraphics[width=0.45\textwidth]{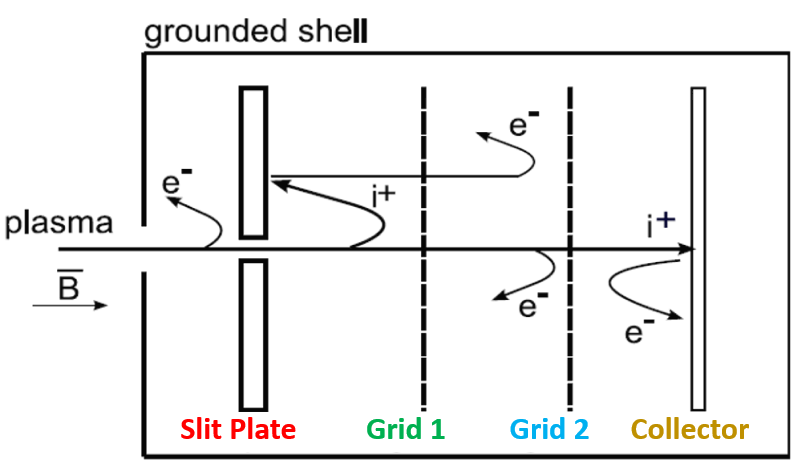}
\caption{Schematic of the MAST-U RFEA module showing the function of the slit plate, grids
and collector plate.\cite{damizia2024first}}
\label{fig:RFEA_Grids}
\end{figure}

\begin{figure}[H]
\centering
\includegraphics[width=0.45\textwidth]{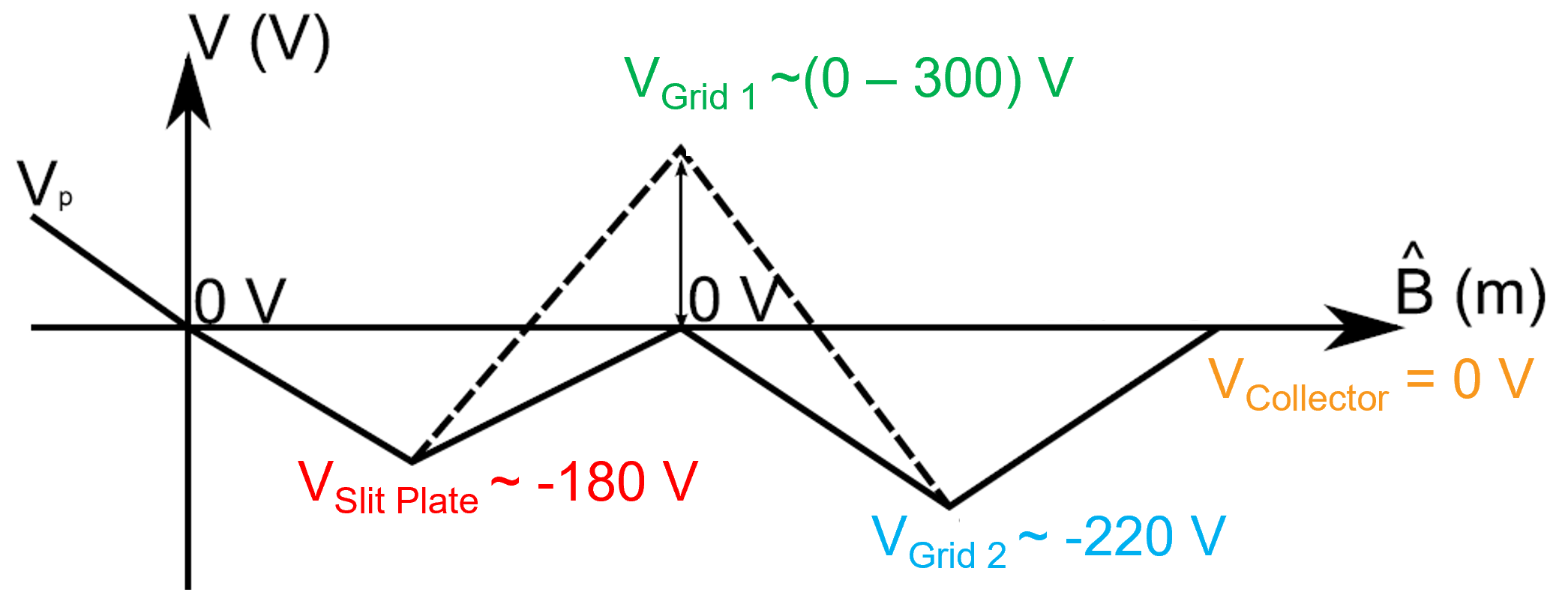}
\caption{Draw of the example voltage settings for each grid of the RFEA.}
\label{fig:RFEAVoltage}
\end{figure}

An RFEA probe consists of a series of grids followed by a collector plate as shown in \autoref{fig:RFEA_Grids} and \autoref{fig:RFEAVoltage}. An RFEA measures the component of the ion velocity distribution perpendicular to the plane of the grid faces. The entrance slit must be wide enough to permit adequate flux transmission,
but also sufficiently small such that the electrostatic sheath established
around the slit edges can shield the aperture from the plasma\cite{pitts1991ion}. This
implies slit widths of the order of the Debye length, which, in tokamak edge plasmas, is usually of the order of tens of microns.

When the particles are transmitted through the aperture in the RFEA, they experience an electric field which varies according to the distance they have traveled into the probe. This electric field is established through bias potentials applied to a number of grids. As the retarding potential is varied, ions of different energies are reflected, or transmitted.
Moreover, the grid can be appropriately biased to reject the species not required; this means that, the probe is able to sample ions or electron accordingly with the analysis mode chosen.
The internal structure of the MAST-U RFEA (\autoref{fig:RFEA_Grids}) and the role of each grid operating in ion analysis mode is described below:

\begin{itemize}
    \item \textbf{Slit plate grid:} This grid is typically maintained at a negative voltage around -180 to repel electrons $V$ and it is in direct contact with a protective slit plate with a fissure length of $5 \, \text{mm}$ and a width of $20 \, \mu \text{m}$. This grid is generally referred to as the plasma electron suppression grid.
    \item \textbf{Grid 1:} The discriminator grid is typically swept from 0 to 120V, retarding the ions to control their passage based on energy. This sweeping allows the probe to selectively measure ions with specific energy levels. Grid 1 can be swept at a selectable rate to optimize measurement conditions.
    \item \textbf{Grid 2:} This grid is held at a constant negative potential to suppress secondary electrons either emitted from the collector, or from the rear of the slit plate due to ion impact.
    \item \textbf{Collector:} The collector plate at the back of the analyzer is made of copper to ensure good conductivity and measures the ions current \(I_{\text{coll}}\).
\end{itemize}

\begin{figure}[H]
\centering
\includegraphics[width=0.48\textwidth]{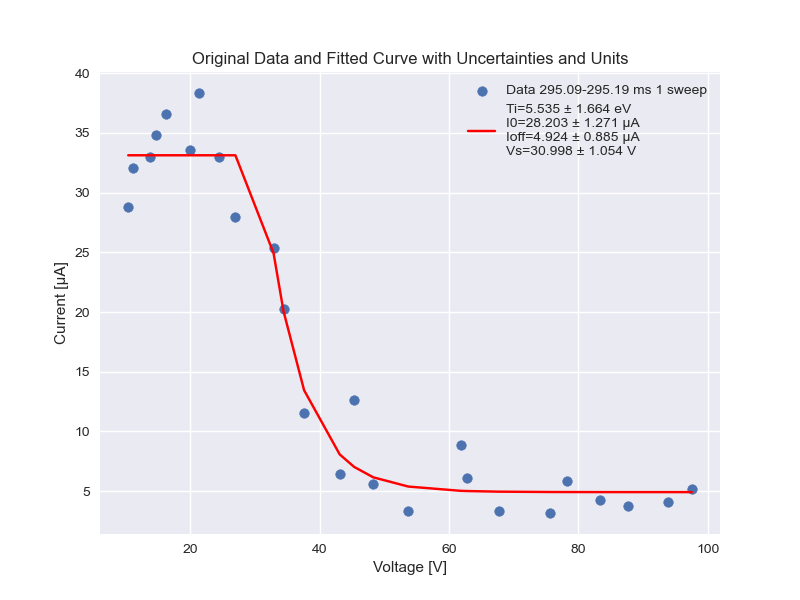}
\caption{ 
Raw data points (blue) of ion current versus applied voltage, gathered from a single sweep between 295.09 - 295.19 ms during plasma shot 47775. These data points have been fitted (red curve) to a theoretical model (Eq: \ref{eq:Fit}).}
\label{fig:Fit_example}
\end{figure}

Assuming that the energy distribution of the ions along the magnetic field lines is described by a Maxwellian distribution, the effective ion temperature $T_{i}$
can be determined by fitting a graph of $I_{col}$ versus grid 1 voltage
($V_{Grid 1}$) using equation (\ref{eq:Fit}):

\begin{equation} 
\label{eq:Fit} 
I_{\text{col}} = I_{0} e^{\left(-\frac{V_{\text{Grid 1}} - V_{s}}{T_{i}}\right)} + I_{\text{off}}
\end{equation}

where $I_{0}$ is the ion saturation current, $V_{s}$ is the plasma sheath voltage and $I_{off}$ is an offset current. The plasma sheath potential in a RFEA is the electric potential difference that forms near the probe’s entrance as a result of the interaction between the plasma and the probe's biased surface. This sheath acts as a potential barrier, which incoming ions must overcome to enter the RFEA. A typical ion current graph is shown in \autoref{fig:Fit_example}.

\subsection{Elongated Divertor Measurements}

\begin{table*}[ht]
\centering
\caption{Summary of Shots}
\label{tab:tokamak_shots}
\begin{tabular}{|c|c|c|c|c|c|}
\hline
\textbf{Shot} & \textbf{$I_p$} [kA] & \textbf{$P_{\text{NBI}}^{tot}$} [MW] & \textbf{time} [ms] & \textbf{$n_e^{\text{core}}$ } [$m^{-3}$] & \textbf{ Configuration} \\ \hline
47775         & 750              & 1.1  & 200 - 800            & $(2.8 - 4.6) \times 10^{19}$ & ED sweeps\\ \hline
48008         & 750              & 1.6  & 450 - 750           & $(1.6 - 3.78) \times 10^{19}$ & ED\\ \hline
\end{tabular}
\end{table*}

\begin{figure}[H]
    \centering
    \includegraphics[width=0.5\textwidth]{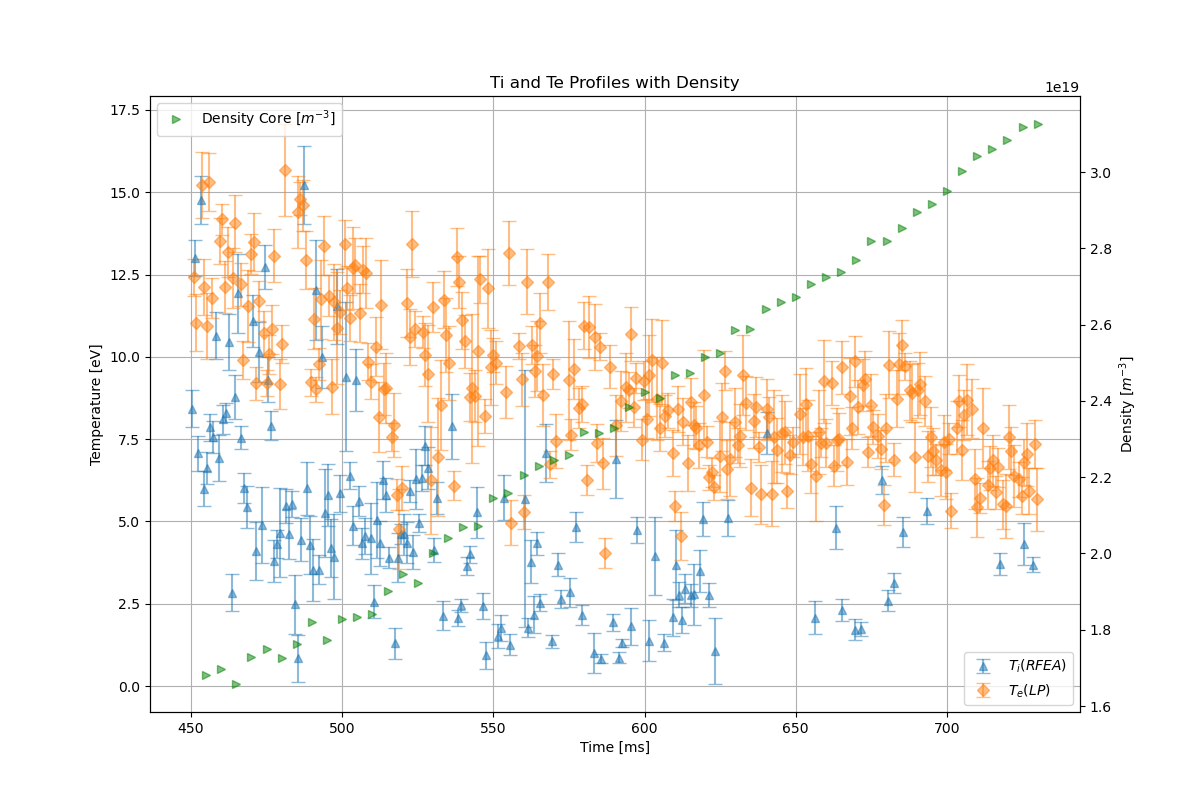}

    \caption[Temporal evolution (\(T_i\)), (\(T_e\)), and core plasma density for Shot 48008]{Temporal evolution of ion temperature (\(T_i\)), electron temperature (\(T_e\)), and core plasma density for Shot 48008.}
    \label{fig:TiTeDensity}
\end{figure}

In this section, the analysis for the temperature profiles of ions (\(T_i\)) and electrons (\(T_e\)) during a steady-state elongated divertor shot 48008 is shown in \autoref{fig:TiTeDensity}, during which the strike point remains constant on the probe. The data were collected using the RFEA for ion temperatures and the LPs for electron temperatures. Regarding  the LP, their measurements under detached conditions require careful interpretation to ensure reliability\cite{fevrier2018analysis} and might be overestimated in this analysis. Additionally, line average core density measurements were recorded over time to observe the effect of density ramping. For this shot, the RFEA Grid 1 was sweeping at a rate of 1 KHz and the temperature measurements are averaged over two voltage sweeps of the Grid 1 for each measurement. To enhance the signal-to-noise ratio (SNR) of the collector current, noise was reduced by applying a smoothed average prior to the fitting process. This method involves using a uniform filter to smooth the data which reduces high-frequency noise. 
The plot in \autoref{fig:TiTeDensity} illustrates the temporal evolution of the ion temperature (\(T_i\)) and electron temperature (\(T_e\)) alongside the line-averaged core density whilst the strike point was on the DSF. 

The time period shown in \autoref{fig:TiTeDensity} includes only the measurements taken when the strike point was positioned on the probe in ED configuration that correspond to $\Delta R_{\text{LCFS}} = 0$.

Firstly, the core plasma density exhibits a consistent increase over the duration of the measurement. Starting at approximately \(1.6 \times 10^{19} \, \text{m}^{-3}\), the density ramps up steadily, reaching around \(3.1 \times 10^{19} \, \text{m}^{-3}\) by the end of the observation period.

Regarding the ion temperature (\(T_i\)), measured by the RFEA, a noticeable initial decreasing trend is observed. Beginning around 11 eV, \(T_i\) gradually decreases to approximately 2 eV. This observed inverse relationship between ion temperature and increasing plasma density aligns with theoretical predictions, evident up to approximately 600 ms, beyond which the expected correlation weakens.
The electron temperature (\(T_e\)), measured by the Langmuir Probes, shows a relatively stable decrease with increasing density. Initial values around 12 eV gradually decrease to about 8 eV as the density increases. The more stable profile of \(T_e\) compared to \(T_i\) could indicate that electron thermal equilibrium is maintained more effectively, even with the increased density. The slight decrease in \(T_e\) may still reflect energy losses due to higher collisionality at elevated densities.

When comparing \(T_i\) and \(T_e\), it is noted that initially, \(T_i\) is lower than \(T_e\). Over time, as the density increases, the gap between \(T_i\) and \(T_e\) narrows, highlighting the differing responses of ions and electrons to the density ramp. This difference in temperature evolution may be attributed to the distinct mechanisms of energy dissipation and confinement affecting ions and electrons differently in a denser plasma environment.

\begin{figure}[H]
    \centering
    \includegraphics[width=0.45\textwidth]{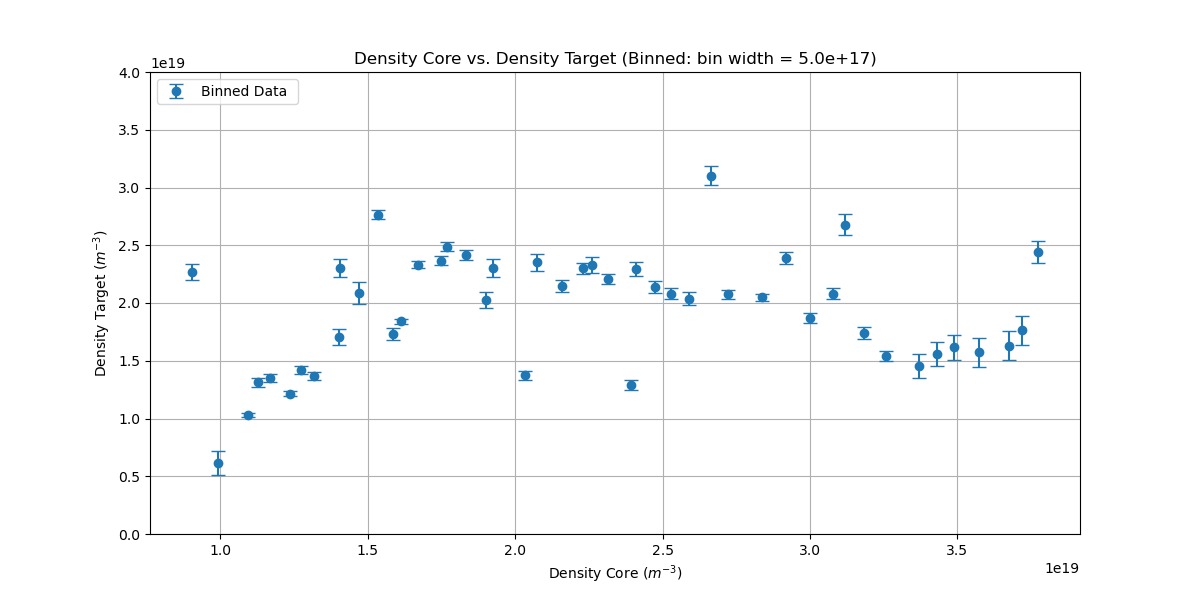}
   \caption[Core vs Target density for shot 48008]{Core vs Target density for shot 48008. The plasma is in ED configuration in the density range between $ 1.6 \times 10^{19} < n_{e}^{core} < 3.2 \times 10^{19}$ $ m^{-3}$.}
    \label{fig:Core_VS_Target_density_48008}
\end{figure}

\autoref{fig:Core_VS_Target_density_48008}, illustrates the relationship between the line average core density and the density at the target, measured by the LP in the divertor region of the MAST-U tokamak. It is possible to distinguish different phases from this plot: Attached Phase, Partial Detachment, and Detached Phase.

\textbf{Attached Phase (Core Density $< 1.6 \times 10^{19} \, \text{m}^{-3}$):} 
In the low-density regime, specifically when the core density is below $1.6 \times 10^{19} \, \text{m}^{-3}$.
During this phase, the strike point is approaching the RFEA location, indicating that the plasma remains connected to the divertor targets. The density at the target shows a rapid increase, which coincides with the ramp-up of core density. To determine the cause of this density rise, we analyzed the timing of both the core density increase and the strike point movement. It appears that the rapid increase in divertor density can be attributed primarily to the approaching strike point, which effectively increases the local plasma pressure in the divertor region. However, the simultaneous increase in core density also contributes to the overall density rise, suggesting a combination of both effects. This dual contribution supports the observation that the plasma remains attached before reaching the probe. 

\textbf{Partial Detachment (Core Density $\approx 1.6 - 2.5 \times 10^{19} , \text{m}^{-3}$):} As the core density exceeds approximately $1.6 \times 10^{19} , \text{m}^{-3}$, the behavior changes, indicating the onset of detachment. In this intermediate phase, the strike point is static over the RFEA and LP on tile 4, allowing for continuous temperature measurements. The density at the target decreases slowly, indicating that the plasma is beginning to detach from the divertor plates. Here, partial detachment means that detachment is in its early stages; the plasma is just starting to separate from the divertor target. This implies that while some detachment is occurring, the divertor is not yet fully detached, and energy and particle fluxes are still partially reaching the target.

\textbf{Detached Phase (Core Density \(> 2.5 \times 10^{19} \, \text{m}^{-3}\)):} 
In the high-density regime, beyond \(2.5 \times 10^{19} \, \text{m}^{-3}\), the plasma shows a clear detachment behavior. The density at the target exhibits a significant rollover, decreasing despite further increases in core density. This detachment signifies that the plasma is no longer fully interacting with the divertor target. Detachment primarily reduces the particle loads to the target. The energy is being radiated away or redistributed, leading to a cooler and less dense plasma at the target.

\begin{figure}[H]
    \centering
    \includegraphics[width=0.45\textwidth]{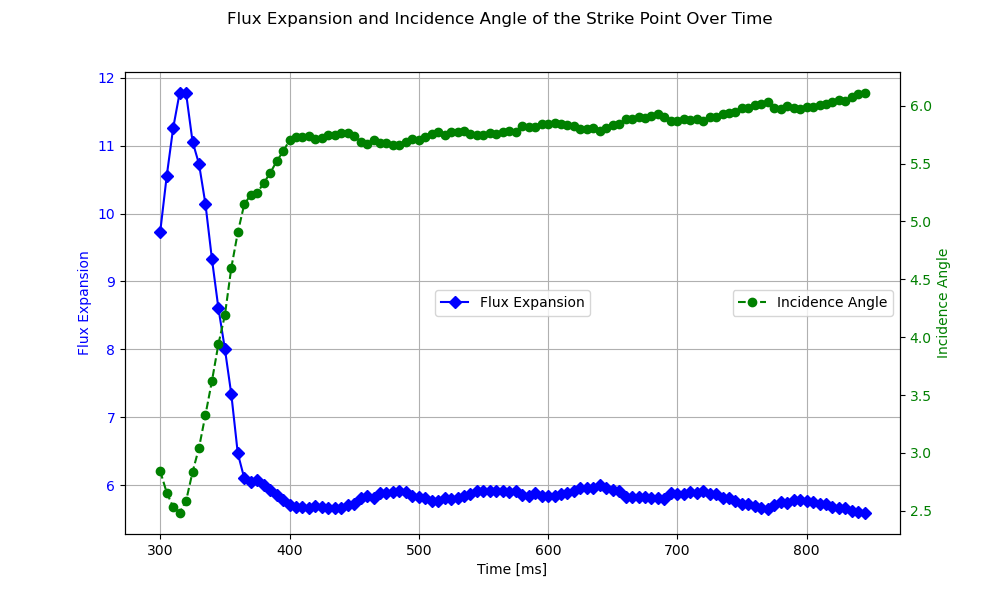}
    \caption{Flux expansion and angle of incidence of the strike point for shot 48008.}
    \label{fig:Flux_Angle_48008}
\end{figure}

\autoref{fig:Flux_Angle_48008} shows the behavior of the poloidal flux expansion and incidence angle over time when the strike point is on the DSF location. The blue diamond illustrates the variation of flux expansion at the target over time.
Flux expansion is defined as the ratio of the flux tube cross-sectional area at the divertor target to its cross-sectional area in the midplane. It characterizes how much the magnetic flux diverges as it moves from the core of the plasma to the divertor region. A higher flux expansion implies that magnetic field lines spread out more widely in the divertor, which helps to reduce the heat and particle flux density reaching the divertor target. Thus, flux expansion is an important parameter as it significantly affects the distribution of heat and particle loads on the divertor target.
Initially, the flux expansion decreases rapidly and then oscillates around an average value of approximately 5.8, indicating low fluctuations in the magnetic field configuration during this period. The initial rapid decrease is indicative of a change in the magnetic equilibrium or plasma shaping, and corresponds in fact to the transition from CD to ED configuration.
The green circles depicts the incidence angle of the strike point on the target over the same time frame. The incidence angle is a critical factor in determining the efficiency and accuracy of measurements taken by the RFEA. The incidence angle starts at around 2.5 degrees in CD, then rises and stabilizes between 5 and 6 degrees for the majority of the measured period in the ED configuration.
The RFEA is aligned at an angle of 5 degrees to the horizontal. The fact that the incidence angle of the strike point remains between 5 and 6 degrees suggests that the measurements are taken under optimal alignment conditions. This near-perfect alignment maximizes the accuracy of the ion temperature measurements by the RFEA, ensuring that the probe is effectively sampling the plasma at the intended orientation.

\subsubsection{Ti/Te Ratio vs. Density}

\begin{figure}[H]
    \centering
    \includegraphics[width=0.45\textwidth]{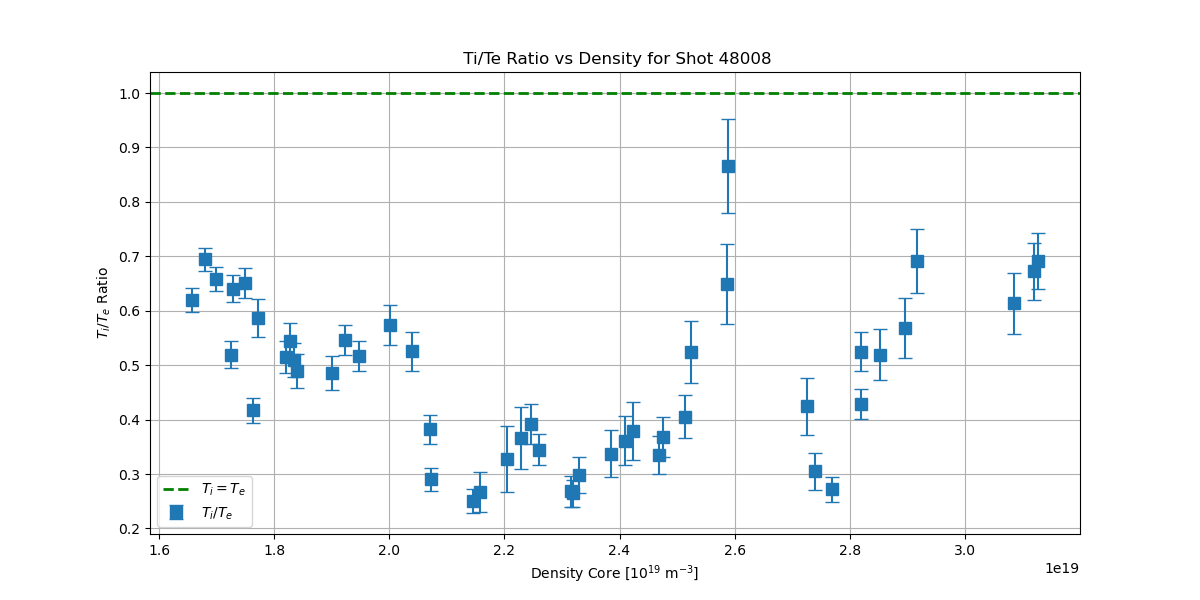}
    \caption[(\(T_i\)) to  (\(T_e\)) ratio as a function of core plasma density for Shot 48008]{Ratio of ion temperature (\(T_i\)) to electron temperature (\(T_e\)) as a function of core plasma density for Shot 48008.}
    \label{fig:TiTeRatioVsDensity}
\end{figure}

Figure \ref{fig:TiTeRatioVsDensity} presents the ratio \(T_i/T_e\) as a function of core plasma density. 

The ratio \( T_i/T_e \) generally decreases with increasing density up to approximately \( 2.4 \times 10^{19} \, \text{m}^{-3} \), which may suggest a more rapid cooling of ions compared to electrons in this range. This observed trend could be attributed to increased ion-neutral collisions, potentially leading to more efficient energy transfer from ions to neutrals. In contrast, electrons might retain more of their energy due to their lower collisionality with neutrals and differing energy loss mechanisms.

Beyond \( 2.4 \times 10^{19} \, \text{m}^{-3} \), the ratio \( T_i/T_e \) appears to increase, indicating a possible stabilization in the relative temperatures as the density continues to rise. This behavior might be associated with changes in energy confinement within the divertor, potentially resulting in reduced cooling of ions compared to the initial detachment phase.

At the highest densities measured (\( 3.1 \times 10^{19} \, \text{m}^{-3} \)), the ratio \( T_i/T_e \) approaches values observed at the lower end of the density range. During this phase, electrons are generally expected to lose energy more efficiently than ions due to radiative losses and collisional processes. However, the more substantial energy loss of ions in the density range \( 2.0 - 2.9 \times 10^{19} \, \text{m}^{-3} \) could be due to enhanced ion-neutral charge exchange and elastic collisions, which may preferentially cool ions.

\begin{figure}[H]
    \centering
    \includegraphics[width=0.48\textwidth]{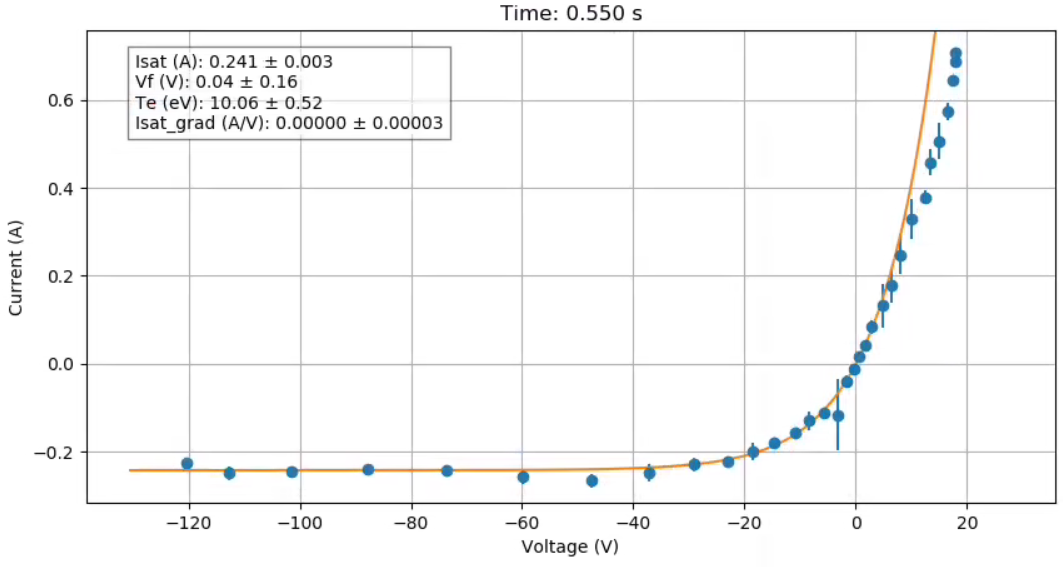}
    \caption{Fitted Langmuir Probe (LP) current-voltage characteristic at \( t = 0.550 \, \text{s} \). The close fit of the data (e.g., \( T_e = 10.06 \pm 0.52 \, \text{eV} \)) suggests high-quality data, yet comparisons with simulations indicate a potential overestimation under detached conditions. This discrepancy underscores the challenges in obtaining accurate electron temperature measurements in detached plasma regimes.}
    \label{fig:LP_fit}
\end{figure}

During the detachment phase, it is likely that the LP overestimates electron temperature (\(T_e\)), an effect attributed to reduced electron saturation current, sheath expansion, and challenges in accurately measuring plasma potential under these conditions\cite{gunn2010retarding}\cite{komm2017effect}. This overestimation could contribute to the observed behavior where the ion-to-electron temperature ratio, \( T_i/T_e \), changes across different density ranges, particularly as full detachment begins.
Simulations, such as those by Rozhansky et al.\cite{rozhansky2013modeling}, predict an electron temperature at the divertor plate of around 4 eV, consistent with typical detached conditions. However, in shot 48008, LP measurements suggest higher values, in the range of 6-8 eV, indicating an overestimation by roughly 1.5 to 2 times the expected \(T_e\). 
This overestimation likely skews the \( T_i/T_e \) ratio calculated from LP data towards lower values than expected. In contrast, the ion temperature (\( T_i \)) measured by the RFEA more closely follows trends predicted by simulations, suggesting that LP-derived \( T_e \) measurements are less reliable under detached conditions. Additionally, the impact of circuit resistance from the plasma on \( T_i \) measurements is mitigated by the greater effective probe sheath resistance, rendering RFEA measurements less susceptible to such effects.
Correcting for this overestimation would yield a \(T_i/T_e\) ratio closer to or above unity, aligning with physical expectations in detached plasmas where ion and electron temperatures are anticipated to equilibrate or where \(T_i\) may exceed \(T_e\) due to enhanced energy exchange.
In summary, the LP's apparent overestimation of \(T_e\) during detachment likely contributes to the anomalously low \(T_i/T_e\) ratios. Adjusting for this effect would make temperature profiles and \(T_i/T_e\) ratios from electrical probes more consistent with simulation predictions and theoretical expectations for detached plasma conditions. 
This discrepancy is visually illustrated in \autoref{fig:LP_fit}, where the fitted Langmuir Probe data, analyzed using a four-parameter model\cite{ryan2023overview}, shows a good match to the current-voltage characteristic, yet indicates higher \( T_e \) values than simulations predict under detachment.

\subsection{ELM Burn Through}

\begin{figure}[H]
\centering
\includegraphics[width=0.50\textwidth]{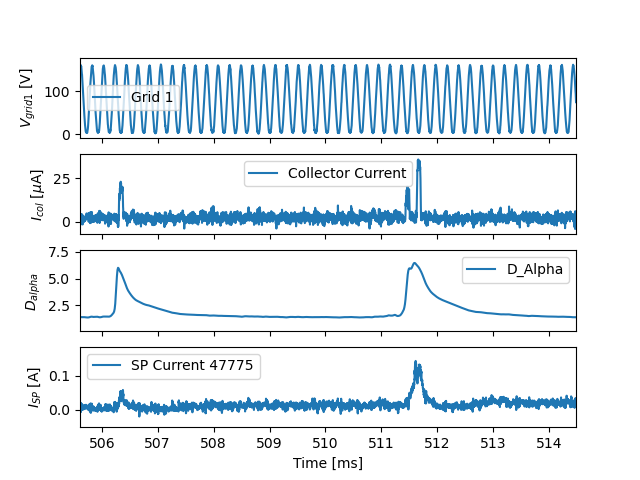}
\caption{Grid 1 voltage (\(V_{\text{grid1}}\)), collector current (\(I_{\text{col}}\)), D\(_\alpha\) signal, and slit plate current (\(I_{\text{SP}}\)) as a function of time for shot 47775, highlighting both ELM and inter-ELM periods.}
\label{Example_Signal_47775}
\end{figure}

For ELM burn-through studies, data from shot 47775 (described in \autoref{tab:tokamak_shots}) was used, where the strike point was repeatedly sweeping back and forth over the RFEA location. This sweeping allowed for multiple ELM burn-through events to be captured. For this shot, the sweep rate of the RFEA $Grid_{1}$ was set at 10.24 kHz, allowing Ti measurements to be obtained every 100 microseconds. Unlike the previous method used in shot 48008, no averaging between sweeps has been applied; instead, a $T_{i}$ measurement is taken for each rising and falling sweep to capture as many points as possible from an ELM.
The probe captured signals only during these events, which were associated with peaks in the D-alpha emission, as shown by the collector current and slit plate current signals (\autoref{Example_Signal_47775}). The measured signals correspond to ELM burn-through events, where the ELMs passed through the detachment layer and reached the probe surfaces. This behavior suggests that the plasma was in a detached state, and only the high-energy ELMs had enough energy to reach the probe through the neutral gas layer formed in the divertor.
This high-frequency approach allows for sufficiently rapid measurements to capture signals from ELMs and to reconstruct temperature profiles using data from multiple ELM events.
The data from the RFEA indicates the presence of ELM burn-through events, as evidenced by the ion current peaks on the collector of the RFEA during ELM phases. In contrast, the inter-ELM phases show no significant signal, highlighting the transient nature of these events ( \autoref{Example_Signal_47775}).

\begin{figure}[H]
\centering
\includegraphics[width=0.48\textwidth]{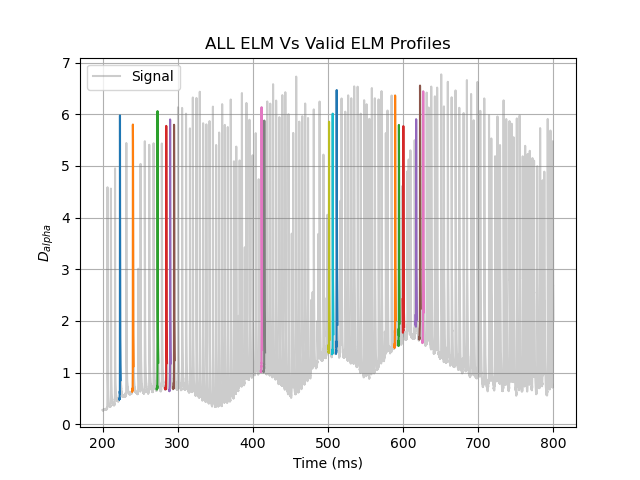}
\caption{This plot displays all detected ELM events from the D\(_\alpha\) signal over time (shown in grey). The colored ELMs represent those that provided a valid signal on the RFEA probe and were utilized for this analysis.
}

\label{Good_ELM_47775}
\end{figure}

To identify and analyze valid ELM events, a custom code method was developed. This method first detects the peaks from the D\(_\alpha\) signal. It then selects a specific range around the D\(_\alpha\) peak and checks the slit plate current for a valid signal. Signals from slit plates that are saturated are excluded from the analysis. Once only valid slit plate signals remain, the method checks the collector signal. The combination of the collector and grid 1 signals is then used to extract the ion temperature. Only those ELM events that produced a clear and measurable signal on the RFEA were selected, ensuring the feasibility of ion temperature extraction. 
\autoref{Good_ELM_47775}, shows the time evolution of \( D_{\alpha} \) temporal evolution during selected ELM events within the time frame of 200 to 800 ms. Each colored line represents a distinct ELM event that was deemed valid for ion temperature analysis.

From the plot it is possible to observe the temporal distribution of the ELM events over the entire shot time frame, with multiple occurrences clustered within specific time intervals. This clustering indicates the moment in which the RFEA has measured an analyzable signal. Signal was not measured when the slit plate power supply saturated or when the collector current signal was below the background signal noise level.

The \( D_{\alpha} \) profiles in \autoref{Good_ELM_47775}, show consistent behavior across different ELM events, with distinct peaks corresponding to the ELM bursts. The consistency of these profiles is crucial for ensuring the reliability of ion temperature measurements derived from the RFEA signals.

\begin{figure}[H]
\centering
\includegraphics[width=0.48\textwidth]{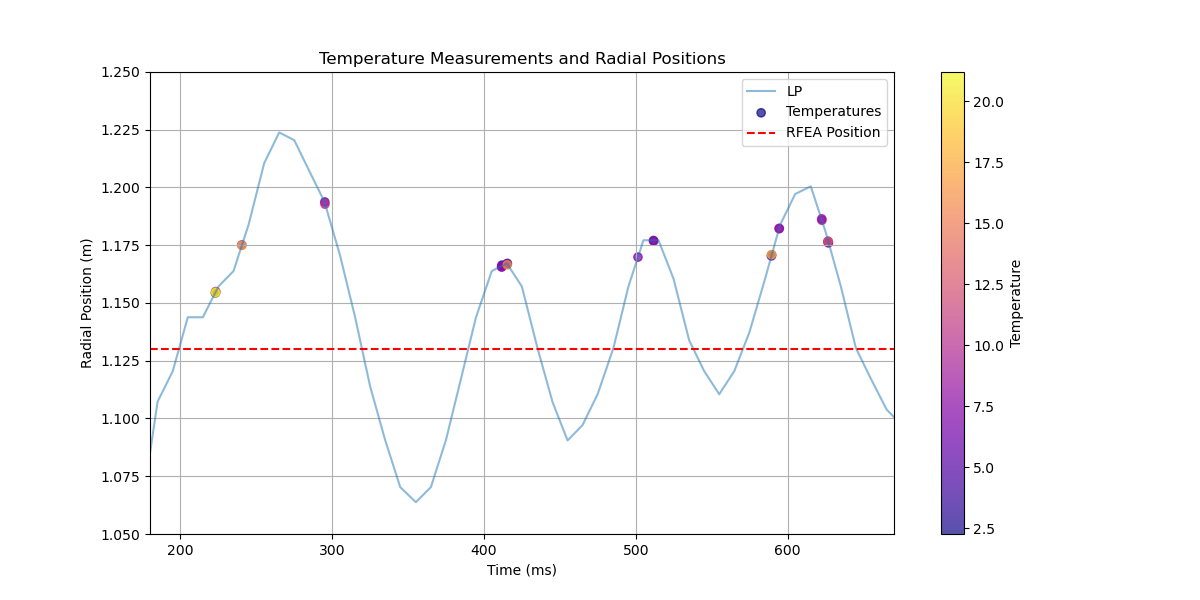}
\caption{Strike point position (blue) from the LP data. The symbols represent the positions where valid $T_{i}$ measurements have been taken from the RFEA. The red dashed line shows the RFEA position.}
\label{Radial_T_47775}
\end{figure}

\autoref{Radial_T_47775}, shows the position of the strike point of the plasma as measured by the LP $J_{sat}$ peak data over time.
The red dashed line indicates the geometrical position of the RFEA, and the blue line represents the strike point location from LP data. The temperature measurements are denoted by colored markers, with the color scale indicating the temperature values. 
The analysis of this plot reveals that the cluster of ELMs shown in \autoref{Good_ELM_47775}, corresponds to different radial positions where the RFEA was able to measure the ion temperature.
By comparing \autoref{Good_ELM_47775} and \autoref{Radial_T_47775}, it becomes evident that only at certain radial positions does the RFEA provided valid measurements during the shot 47775. In \autoref{Radial_T_47775}, the points where the temperature measurements were taken represent data collected in the private flux region (PFR), as the strike point was located beyond the RFEA at 1.13 m. There are no reliable measurements at the strike point or in the SOL because the slit plate was saturating, preventing the probe from collecting any usable signal at those stages.

\begin{figure}[H]
\centering
\includegraphics[width=0.48\textwidth]{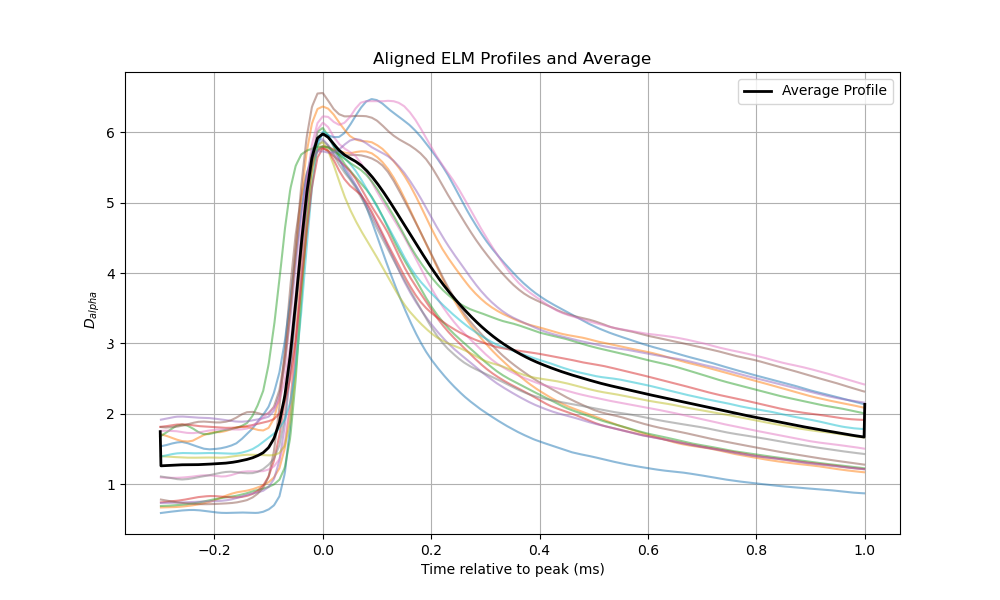}
\caption{ $D\alpha$ signal traces of the valid ELMs measured from the RFEA as a function of time relative to the peak in the $D\alpha$ trace. A mean of the profiles is shown in black.}
\label{ELM_average_47775}
\end{figure}

\autoref{ELM_average_47775} illustrates the average $D_{\alpha}$ signal during ELMs for shot 47775. This profile was constructed by averaging the $D_{\alpha}$ signals from all ELM events that met the criteria for valid analysis. The resulting profile provides a representative view of the typical $D_{\alpha}$ signal behavior associated with ELM events in this shot.

\begin{figure}[H]
\centering
\includegraphics[width=0.48\textwidth]{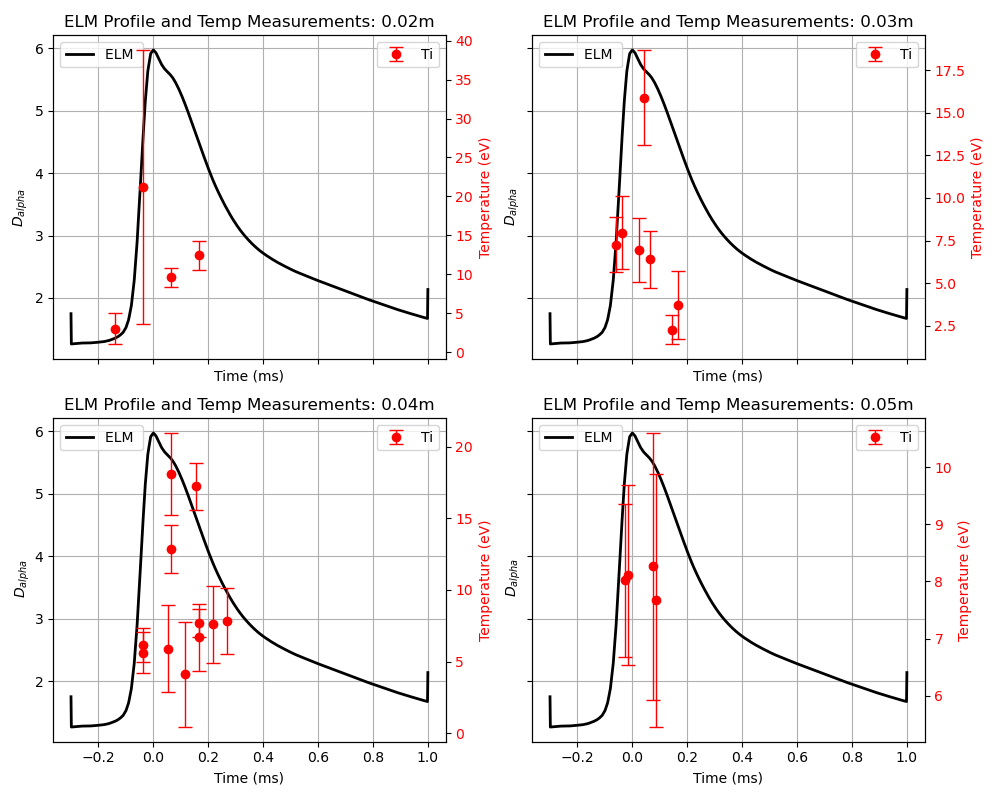}
\caption{Composite of ion temperatures measured during ELMs by the RFEA as a function of distance from the RFEA. The black profile shows an average ELM $D_{\alpha}$ profile for illustrative purposes.}
\label{Radial_Band_Ti_47775}
\end{figure}

In figure \ref{Radial_Band_Ti_47775} multiple subplots present the ion temperature measurements, at different distances from the RFEA during the strike point sweep. Each subplot corresponds to a specific radial distance, as calculated from the strike point data obtained from the LPs. The x-axis in each subplot represents time relative to the ELM peak, and the y-axis shows both the \( D_{\alpha} \) signal on the left, and the ion temperature (\( T_i \)) on the right.

From these subplots, it is evident that the ion temperature measurements vary with radial distance from the RFEA. The temperature peaks around 22 eV at 0.02m from the RFEA. Notably, the density of valid measurements appears to correlate with certain radial positions. Specifically, where more valid measurements are observed, it might indicate that the magnetic field is well-aligned with the grids of the RFEA or the strike point was near, optimizing the measurement conditions. Furthermore the saturation of the power supply has affected numerous ELM measurement reducing the statistics available for this analysis. 
Alternatively, this pattern could indicate a discrepancy between the actual position of the probe and the radial position of the strike point as measured by the LP, potentially due to measurement errors, slight misalignments, or more likely, toroidal asymmetry. Such asymmetry can cause variations in the strike point position around the torus, leading to inconsistencies between measurements taken at different toroidal locations\cite{damizia2024first}.

These observations underscore the importance of precise alignment and accurate positioning in obtaining reliable ion temperature measurements with an RFEA, particularly during transient events where saturation can occur. To optimize the measurements, increasing the power supply threshold is recommended. The current power supply saturates at 3 Amperes in this case; therefore, a higher tolerance would enable the measurement of high-energy ELMs and allow for the reconstruction of complete profiles in both the SOL and PFR regions.

\section{Conclusion}

The experimental analysis conducted using the RFEA on the MAST-U divertor has provided valuable insights into the behavior of ion temperatures under different plasma conditions, particularly in the elongated divertor and during ELM burn-through events.

In the elongated divertor configuration (Shot 48008), the data indicated distinct phases of plasma attachment and detachment as core density increased. Initially, during the attached phase, the plasma appeared well-connected to the divertor targets, with both ion and electron temperatures decreasing with increasing density.

The analysis of ion and electron temperatures showed that the ion temperature (\(T_i\)) decreased more sharply with increasing density compared to the electron temperature (\(T_e\)). 
In the detached phase, the electron temperature profile, as measured by the Langmuir probes, flattened around 8\,eV; however, it is important to note that LP measurements are often unreliable under detached conditions. Consequently, the ion temperature profile became more scattered around 2-5 eV, reflecting the challenges in obtaining accurate measurements during detachment.
The \( T_i/T_e \) ratio provided additional insights, indicating a complex interplay of thermal dynamics, particularly in high-density plasmas. During the analysis of this shot, the ratio \( T_i/T_e \) consistently remained below one. In contrast, in previous experiments on MAST, the \( T_i/T_e \) ratio was typically observed to be higher than one. This difference may be attributed to the elongated divertor configuration in MAST-U, which likely enhances plasma exhaust and promotes detachment.
In the ELM burn-through analysis (Shot 47775), the RFEA successfully captured the transient nature of ELM events, showing significant ion current peaks during ELM phases and negligible signals during inter-ELM periods. The radial position analysis highlighted the importance of precise alignment between the strike point and the RFEA for obtaining accurate measurements. The findings from this shot suggest the efficacy of the RFEA in detecting ELM burn-through phenomena and provide valuable data on the interaction between ELMs and the divertor region. For both shots, further experiments similar to these are required in the next campaign to study the evolution of the \(T_i\) measurements obtained from the RFEA. 
The objective is to optimize the grid settings by lowering the slit plate voltage to a higher negative value, upgrading the power supply to increase the saturation threshold, and modifying other grid settings. Additionally, different plasma conditions and levels of detachment will be explored. The use of a cryogenic pump and a higher power heating system in the MU04 campaign is expected to facilitate more attached plasma conditions. The measurements presented here are from the initial data obtained during the first installation of the RFEA in the DSF during the MU03 campaign. Further experiments are planned for the MU04 campaign to expand on these findings and improve our understanding.

\section*{Acknowledgements}
\fontsize{8pt}{8pt}\selectfont
This work has been funded by the EPSRC Energy Programme, grant EP/S022430/1, EP/T012250/1, EP/N023846/1, EP/W006839/1 and the University of Liverpool. This work has been carried out within the framework of the EUROfusion Consortium, partially funded by the European Union via the Euratom Research and Training Programme (Grant Agreement No 101052200 — EUROfusion). 
This work is supported by US Department of Energy, Office of Fusion Energy Sciences under the Spherical Tokamak program, contract DE-AC05-00OR22725. 
Views and opinions expressed are however those of the author(s) only and do not necessarily reflect those of the European Union or the European Commission. Neither the European Union nor the European Commission can be held responsible for them. For the purpose of open access, the author(s) has applied a Creative Commons Attribution (CC BY) licence (where permitted by UKRI, ‘Open Government Licence’ or ‘Creative Commons Attribution No-derivatives (CC BY-ND) licence’ may be stated instead) to any Author Accepted Manuscript version arising.

\end{multicols}
\end{document}